\documentclass[12pt,showpacs,preprintnumbers,amsmath,amssymb]{revtex4}
\usepackage{dcolumn}
\usepackage{epsfig}
\usepackage{psfrag}

\newcommand{\be}{\begin{equation}}
\newcommand{\ee}{\end{equation}}
\newcommand{\bea}{\begin{eqnarray}}
\newcommand{\eea}{\end{eqnarray}}

\newcommand{\WW}{{\rm W}}
\newcommand{\F}{{Y}}
\newcommand{\A}{{Y}}
\newcommand{\thetw}{\theta_{\mbox{{\tiny W}}}}
\newcommand{\mh}{m_{\mbox{{\tiny H}}}}
\newcommand{\mz}{m_{\mbox{{\tiny Z}}}}
\newcommand{\mw}{m_{\mbox{{\tiny W}}}}

\newcommand{\XPEH}{\it}
\newcommand{\bbf}{\bf}
\begin{document}

\title{Spinning Electroweak Sphalerons}
\author{Eugen Radu and Mikhail S. Volkov}
\affiliation{LMPT CNRS-UMR 6083,
Universit\'e de Tours, Parc de Grandmont,
37200 Tours, FRANCE}

\begin{abstract}
 {We present numerical evidence for the existence of 
stationary spinning generalizations for 
the static sphaleron in the Weinberg-Salam theory. 
Our results suggest that, 
 for any value of the mixing angle $\thetw$
and for any Higgs mass, the spinning sphalerons 
comprise a family labeled by their 
angular momentum $J$. For $\thetw\neq 0$ they possess 
 an electric charge $Q=eJ$ where $e$ is the electron charge.  
Inside they contain a monopole-antimonopole pair and a spinning 
loop of electric current, and for 
large $J$ they show a Regge-type behavior. It is likely that 
these sphalerons mediate the topological transitions in sectors with
$J\neq 0$, thus enlarging the number of transition channels. 
Their action {\it decreases} with $J$, 
which may considerably 
affect the total transition rate. 
}
\end{abstract}
\maketitle

\noindent
\section{Introduction}
The sphaleron represents one of the best known examples 
of solitons in the electroweak sector of the standard model. 
This is a static, purely magnetic solution of the classical field equations 
describing a localized, globally regular object with a finite mass of order of 
several TeV \cite{KM}. 
For vanishing mixing angle $\thetw$ sphaleron is spherically symmetric 
\cite{KM}, but for 
$\thetw\neq 0$ it is only axially symmetric -- due to a nonzero
magnetic dipole moment \cite{KK}.  
The sphaleron is unstable and relates to the potential barrier between 
the topological vacua in the theory \cite{KM}, thereby  mediating 
nonperturbative transition processes that could be relevant
for the generation of the baryon number asymmetry of our Universe \cite{RS}. 

In this article we show that the static, purely magnetic sphaleron 
admits stationary generalizations including an electric field and supporting
a nonzero angular momentum $J$.  This reveals  new 
solitonic states and also provides the first
explicit example of stationary {\it spinning} solitons in the standard model. 
In fact, it seems
that spinning systems in classical field theory should generically radiate and therefore 
cannot be stationary, while spinning not accompanied by radiation 
should be viewed as something exceptional \cite{RV}.  For example,  the existence of
{stationary} spinning generalizations can be ruled out for the magnetic monopoles
\cite{VR}, \cite{VW}.  A similar no-go statement can also be proven for the
sphalerons, but only assuming that they do not depend explicitly on time \cite{VW}, which
is not the case for our solutions below.      

The inner structure of the spinning sphaleron shows a monopole-antimonopole pair 
joined by a Z-string segment and surrounded by a loop of electric current. 
The momentum circulating along the loops gives rise to the angular momentum, 
which can be regarded as an electroweak analogue of cosmic 
vortons \cite{DS}, \cite{RV}. For large $J$  the whole system shows a Regge-type behavior, 
similar to what  was suggested long ago by Nambu \cite{Nambu}.

For  $\thetw\neq 0$ the spinning sphalerons carry an electric charge $Q=Je$ where
$e$ is the electron charge.  
Since $J\in\mathbb{Z}$  in the full quantum theory, it follows that 
only solutions with $Q/e\in\mathbb{Z}$
are allowed. The charged, spinning sphalerons comprise therefore a discrete set. 
It is likely that 
they mediate the topological transitions in sectors with 
fixed charge and angular momentum,
thus enlarging the number of transition channels.  Their energy increases but the 
action  {\it decreases} with $J$, which  
may considerably affect the total transition rate and thus be 
important for the theory of baryogenesis.

\noindent
\section{Weinberg-Salam theory}
Its bosonic sector is described 
by the action $S=\frac{1}{{\rm g}_{z}^2}\int{\cal L}\, d^4x$ where 
\bea                             \label{WS}
{\cal L}&=&
-\frac{1}{4g^2}\,\WW^{\rm a}_{\mu\nu}\WW^{{\rm a}\mu\nu}
-\frac{1}{4g^{\prime 2}}\,{\F}_{\mu\nu}{\F}^{\mu\nu} 
+(D_\mu\Phi)^\dagger D^\mu\Phi
-\frac{\beta}{8}\left(\Phi^\dagger\Phi-1\right)^2.
\eea
Here 
$
\WW^{\rm a}_{\mu\nu}=\partial_\mu\WW^{\rm a}_\nu
-\partial_\nu \WW^{\rm a}_\mu
+\epsilon_{{\rm abc}}\WW^{\rm b}_\mu\WW^{\rm c}_\nu
$ 
and 
$
{\F}_{\mu\nu}=\partial_\mu{\A}_\nu
-\partial_\nu{\A}_\mu
$ 
while
$\Phi$ 
is a doublet of complex Higgs fields with 
$
D_\mu\Phi
=\left(\partial_\mu-\frac{i}{2}\,{\A}_\mu
-\frac{i}{2}\,\tau_{\rm a} \WW^{\rm a}_\mu\right)\Phi
$ where $\tau_{\rm a}$ are the Pauli matrices. 
All fields and spacetime coordinates 
have been rendered dimensionless by rescaling. The 
rescaled gauge couplings are
expressed in terms of the Weinberg angle as 
$g=\cos\thetw$, $g^\prime=\sin\thetw$. 
The mass scale is ${\rm g}_z\Phi_0$ where 
$\Phi_0$ is the dimensionfull 
Higgs field vacuum expectation value, 
the electron  charge is $e={\rm g}_z gg'$.
The theory is invariant under   
gauge transformations
\be                               \label{gauge}
\Phi\to {\rm U}\Phi,~~~~~~~~
{\cal W}_\mu\to {\rm U}({\cal W}_\mu
+2i\partial_\mu){\rm U}^{-1}\,,
\ee
where ${\rm U}\in{\rm SU}(2)\times{\rm U}(1)$ and 
$
{\cal W}_\mu=
\A_\mu+\tau^{\rm a}\WW^{\rm a}_\mu. 
$
The electromagnetic field can be defined in a gauge invariant way as 
$
F_{\mu\nu}=\frac{g}{g^\prime}\,Y_{\mu\nu}-
\frac{g^\prime}{g}\,n_{\rm a}\WW^{\rm a}_{\mu\nu}\,
$
with $n_{\rm a}=(\Phi^\dagger\tau_{\rm a}\Phi)/(\Phi^\dagger\Phi)$ \cite{Nambu}.
The electric current is $j_\mu=\partial^\nu F_{\nu\mu}$
while the dual of $F_{\mu\nu}$ determines similarly  the magnetic current. 

\noindent
\section{Axial symmetry}
Let us split the spacetime coordinates as $x^k=(\rho,z)$ 
with $k=1,2$ and 
$x^a=(t,\varphi)$. 
We are interested in stationary, axially symmetric systems for which
$
K_{(a)}=\partial/{\partial x^a}
$
are the symmetry generators. The existence of these symmetries implies 
conservation of two Noether charges
$
\int T^0_\mu K^\mu_{(a)} d^3{\bf x}
$
which are, respectively, the energy $E$ and angular momentum $J$ for $a=0,3$ 
(the dimensionfull  values being 
$\Phi_0 E/{\rm g}_z$ and $J/{\rm g}_z^2$). 
The energy-momentum tensor is obtained by varying the Lagrangian \eqref{WS}
with respect to the spacetime metric, 
$
T^\mu_{\nu}=2g^{\mu\sigma}{\partial\mathcal{L}}/
{\partial g^{\sigma\nu}}
-\delta^\mu_{\nu}\mathcal{L}\,. 
$

The two Killing vectors $K_a$ commute between themselves. 
Since all the internal 
symmetries in the theory  \eqref{WS} are gauged, there exists 
a gauge where the symmetric fields do not depend on $x^a$ \cite{FM}. 
The {most general} stationary and axially symmetric fields can therefore 
be chosen in the form 
$\Phi=\Phi(x^k)$, ${\cal W}_\mu={\cal W}_\mu(x^k)$.
These  can then be consistently truncated by imposing 
the on-shell conditions
$\Im(\Phi)=0$, $\WW^2_a=\WW^1_k=\WW^3_k=\A_k=0$, 
such that the fields can be parametrized as
\be                              \label{ax}
{\cal W}=(\A_a+\tau_{\rm 1}\psi^{1}_a+\tau_{\rm 3}\psi^{3}_a)dx^a+\tau_2 
v_k dx^k,~
\Phi=
\left[\begin{array}{c}
\phi_{+} \\
\phi_{-}
\end{array}\right]. 
\ee
This is in fact a version of the Rebbi-Rossi ansatz \cite{RR}. 
This parametrization defines the  
reduced field theory for the complex scalars 
$\psi_a=\psi^{1}_a+i\psi_a^{3}$ and $\phi=\phi_{+}+i\phi_{-}$ and a vector 
$v_k$ 
living on the 2D space spanned by $x^k$. 
The field equations following from \eqref{WS},\eqref{ax}
 read 
\begin{subequations}                \label{eqs}
\begin{align}
&\frac{1}{\rho}\,\partial_k\left(\rho h^{ab}\,\partial_k \A_b\right)
=
2{g^{\prime 2}}\Im({\phi}{\lambda^{a}}),            \label{eqs4} \\
&\frac{1}{\rho}\,D_k\left(\rho\,h^{ab}\,D_k \psi_b\right)
=
2{g^{2}}{\phi}^\ast{\lambda^a}
+\frac12(\epsilon^{cd}\psi_c\psi_d^{\ast})\epsilon^{ab}\psi_b,     \label{eqs2} \\
&\frac{1}{\rho}\,\partial_s\left(\rho V_{sk}\right)
=
\Im\{\psi^{a\ast} D_k\psi_a+g^2\phi^\ast{\cal D}_k\phi\}, \label{eqs3} \\
&\frac{1}{{\rho}}\,{\cal D}_k\left({\rho}\,{\cal D}_k 
\phi\right)+{\psi^\ast_a}\lambda^a+i{\lambda^\ast_a}\A^a
=\frac{\beta}{4}(|\phi|^2-1)\phi .  \label{eqs1} 
\end{align}
\end{subequations}                
Here 
$D_k=\partial_k-iv_k$, 
${\cal D}_k=\partial_k+\frac{i}{2}v_k$ 
and $\epsilon^{03}=-\epsilon^{30}=1/\rho$ 
with 
$
\lambda_a=\frac14(\phi\psi_a+i{\phi}^\ast\A_a)
$
also $V_{ik}=\partial_iv_k-\partial_k v_i$.
The
indices $a,b$ are
raised and lowered by the 
`target space' metric
$h_{ab}={\rm diag}(1,-\rho^2)$.
The residual symmetry of the ansatz \eqref{ax} 
generated by 
${\rm U}=\exp(\frac{i}{2}\xi\tau_2)$ gives rise to the 
local U(1) symmetry of Eqs.\eqref{eqs},
\be                    \label{U1}
\psi_a\to e^{i\xi}\psi_a,~~
\phi\to e^{-\frac{i}{2}\xi}\phi,~~v_k\to v_k+\partial_k\xi. 
\ee
Modulo this symmetry, zero energy fields 
 are given by 
\be                  \label{vac}
{\cal W}_0=(\tau^3-1)\,(\omega dt+n d\varphi),~~~~ 
\Phi_0=
\left[\begin{array}{c}
1 \\
0
\end{array}\right] ,
\ee
\noindent
with constant $\omega,n$.  
Eqs.\eqref{eqs} also admit a discrete  symmetry under $z\to -z$,
\be                                  \label{plane} 
\A_a\to \A_a,~\psi_a\to-\psi_a^\ast,~~\phi\to\phi^\ast,~~v_k dx^k\to-v_k dx^k.
\ee

\section{Boundary conditions} 
Let $z+i\rho=re^{i\vartheta}$. 
Finite energy fields 
should approach \eqref{vac} for $r\to\infty$, 
so that one should have 
$
{\cal W}={\cal W}_0+\delta{\cal W}$,
$
\Phi=\Phi_0+\delta\Phi$.
Linearizing Eqs.\eqref{eqs} with respect to 
$\delta{\cal W}$, $\delta\Phi$  gives (in the $\delta\phi_{-}=0$ gauge) 
$$
\delta Y_a+\delta\psi^3_a\sim e^{-\mz r},~
\delta\phi_{+}\sim e^{-\mh r},~
\delta\psi^{1}_a\sim v_k\sim e^{-\mw r},
$$
which correspond,  respectively, to
the Z, Higgs and W bosons with masses 
(in units of ${\rm g}_z\Phi_0$)
\be              \label{mass}
\mz=\frac{1}{\sqrt{2}},~~\mh=\sqrt{\beta}\,\mz,~~~\mw=\sqrt{\frac{g^2}{2}-\omega^2}.
\ee 
We see that the W-boson mass gets screened, since having $\omega\neq 0$ 
is equivalent to 
a manifest time-dependence of the fields. 
For the photon field Eqs.\eqref{eqs} give 
for large $r$
\be                                                        \label{Qmu}
(\frac{g}{g^\prime}\,\delta Y_a-\frac{g^\prime}{g}\,\delta\psi_a^3)dx^a=
\frac{Q}{4\pi r}\,dt +\frac{\mu}{4\pi r}\,\sin^2\vartheta d\varphi+\ldots,
\ee
where $Q,\mu$ are the electric charge and magnetic moment.

In the gauge \eqref{ax} the fields depend only on $x^k$ but they 
have a Dirac string singularity
at the $z$-axis. If $n\in\mathbb{Z}$ then this singularity can be removed by 
the gauge transformation 
$
{\rm U}=\mbox{\boldmath $\tau$}_{1}\exp\{\frac{i}{2}\chi(1-\tau_3)\}  
$
with $\chi=\omega t+n\varphi\equiv \omega_a x^a$. 
Applying this to \eqref{ax} gives 
\bea                     \label{fields}
{\cal W}&=&(Y_a+\omega_a+\mbox{\boldmath $\tau$}_{{\rm 1}}K^{1}_{a}
-\mbox{\boldmath $\tau$}_{{\rm 2}}K^2_{a})dx^a  \notag \\
&-&\mbox{\boldmath $\tau$}_{{\rm 3}}v_k dx^k
+2i\mbox{\boldmath $\tau$}_{{\rm 1}}d\mbox{\boldmath $\tau$}_{{\rm 1}} ,~~~
\Phi=
\mbox{\boldmath $\tau$}_{{\rm 1}}\left[\begin{array}{c}
~~~~\phi_+ \\
e^{i\chi}\phi_{-}
\end{array}\right]
\eea
with 
$
\mbox{\boldmath $\tau$}_{1}=\tau_1\sin\vartheta\cos\chi+
\tau_2\sin\vartheta\sin\chi+\tau_3\cos\vartheta
$, $\mbox{\boldmath $\tau$}_{2}=\partial_\vartheta\mbox{\boldmath $\tau$}_{1}$,
$\mbox{\boldmath $\tau$}_{3}=\partial_\chi\mbox{\boldmath $\tau$}_{1}/\sin\vartheta$,
and
$
K^2_a+iK^1_a=
e^{i\vartheta}(\psi_a-i\omega_a). 
$
The  Dirac string is now absent,
but the fields depend explicitly  on $t,\varphi$. The boundary conditions at infinity 
and in the equatorial plane are specified by Eqs.\eqref{vac},\eqref{plane}. 
Transforming \eqref{fields} to Cartesian coordinates one requires that all term proportional 
to $1/\rho$ and $1/r$ should vanish at the $z$-axis and at the origin, respectively.  
At least for  $n=\pm 1$ no additional 
complications at the axis then arise \cite{Kleihaus} and  the fields
\eqref{fields} are everywhere regular.
These boundary conditions still allow for a residual gauge freedom
\eqref{U1} with $\xi$ vanishing for $\rho=0$, for $z=0$, and for $r=\infty$. 
This freedom can be fixed by the gauge condition $\partial_k(\rho v_k)=0$.

\noindent
\section{Angular momentum}
If $g,g^\prime\neq 0$,  then 
using Eqs.\eqref{eqs}
one can represent 
$T^0_\varphi$ 
as
a total derivative \cite{VR}, \cite{VW}
\be
T^0_\varphi=
\frac{n}{gg^\prime}\,\frac{1}{\rho}\,\partial_k(\rho F_{0k})+\ldots 
\ee
the dots denoting the terms that 
vanish  upon integration. As a result, choosing $n=1$,   
\be                                 \label{J}
J=\int T^0_\varphi d^3{\bf x}
=\frac{1}{gg^\prime}\oint \vec{\cal E}d\vec{S}
=\frac{Q}{gg^\prime},
\ee
where $\vec{{\cal E}}$ is the electric field. 
So far we have used the relativistic 
units where 
$\hbar=c=1$, but let us return for a moment to 
the standard units where the electron 
charge is 
$e=c\hbar{\rm g}_zgg^\prime$. 
Dividing Eq.\eqref{J} by $c\hbar {\rm g}_z^2$ gives then 
the relation for the 
dimensionfull quantities, 
\be                           \label{JQ}
J=Q/e
\ee
with $J$ expressed in units of $\hbar$. Since $J\in\mathbb{Z}$ in the full quantum theory,
it follows that only solutions with $Q/e\in\mathbb{Z}$ are allowed. 
The sphaleron charge is therefore quantized and the 
charged, spinning sphalerons comprise a discrete family.

\noindent 
\section{The $\thetw=0$ limit} 
\begin{figure}[h]
\hbox to\linewidth{\hss%
\psfrag{xxx}{{\Large $\ln(1+r)$}}
  \psfrag{H1}{\Huge $\frac{H_{+}}{r\omega}$}
 \psfrag{H3}{\Huge $\frac{H_{-}}{r\omega}$}
  \psfrag{w}{\Huge $w$}
 \psfrag{h}{\Huge $h$}
 \psfrag{bb}{\Large $\beta=1$}
	\resizebox{7cm}{6cm}{\includegraphics{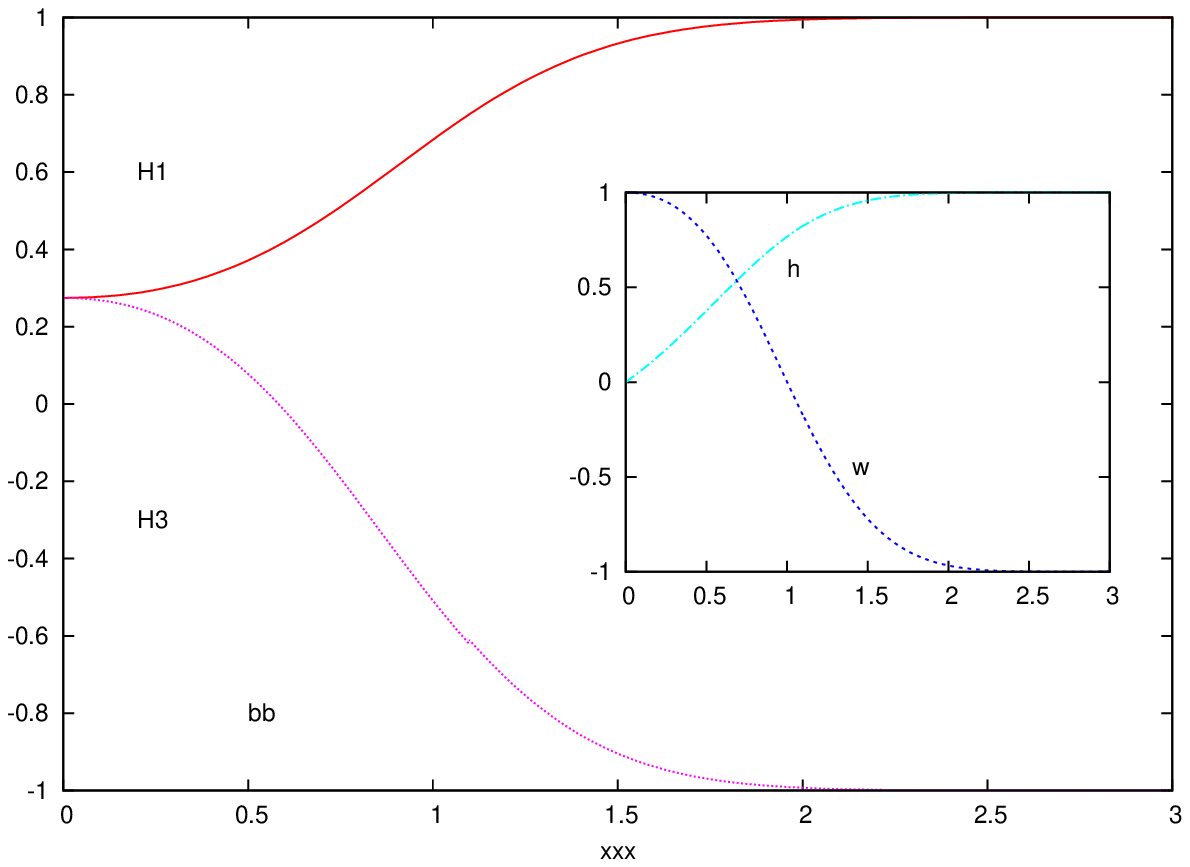}}
\hspace{15mm}%
       \resizebox{7cm}{6cm}{\includegraphics{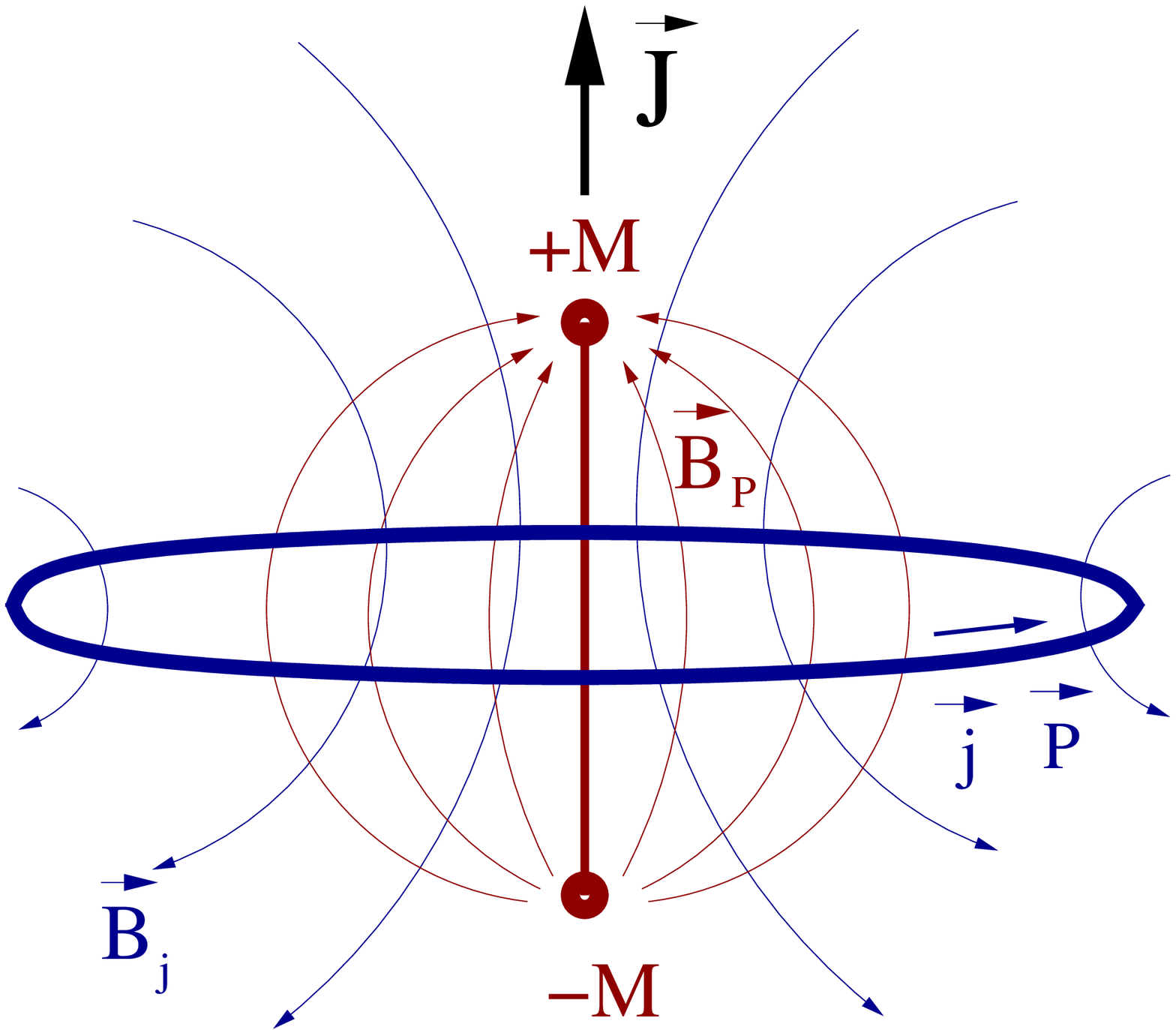}}	
\hss}
\caption{\label{fig1}\small 
Left: perturbative solutions of Eqs.\eqref{p3} and the background sphaleron profiles.
Right: schematic charge-current distribution inside the spinning sphaleron.
}
\end{figure}

When $g=1$ and $g^\prime =0$ the right hand side of 
Eq.\eqref{eqs4} vanishes and so 
the U(1) amplitudes are constant and equal to their asymptotic values, $\A_a=-\omega_a$.
The U(1) part of the gauge field \eqref{fields} then vanishes.
Let us consider the  
spherically symmetric sphaleron \cite{KM},
\be                     \label{KM}
{\cal W}=(w(r)-1)(\mbox{\boldmath $\tau$}_{{\rm 3}}d\vartheta
-\mbox{\boldmath $\tau$}_{{\rm 2}}\sin\vartheta\, d\varphi),~
\Phi=
\mbox{\boldmath $\tau$}_{{\rm 1}}\left[\begin{array}{c}
h(r) \\
0
\end{array}\right],
\ee
which is a particular case of the axially symmetric field \eqref{fields}
with $\omega=0$, $n=1$. Equations \eqref{eqs}  reduce then to
\begin{align}                     \label{spl}
w''&=\frac{w(w^2-1)}{r^2}+\frac12\,h^2(w+1), \notag \\
h''+\frac2r\,h'&=\frac{(w+1)^2}{2r^2}h+\frac{\beta}{4}(h^2-1)h, 
\end{align}
whose globally regular solution exists for any $\beta\geq 0$ \cite{KM} and 
has the profile shown in Fig.\ref{fig1}.

The strategy now is to vary $\omega$ by keeping $n=1$.   
If  $\omega\neq 0$ then the solution  should be sought within the  full ansatz 
\eqref{fields}, but for $|\omega|\ll 1$ it is expected to be close
to the  sphaleron \eqref{KM}. Specifically, in the gauge \eqref{ax}
where the fields do not depend on $t,\varphi$ one should have 
$
{\cal W}={\cal W}_{\rm s}+\delta{\cal W}$,
$\Phi=\Phi_{\rm s}+\delta\Phi$ 
where ${\cal W}_{\rm s},\Phi_{\rm s}$ is the sphaleron fields \eqref{KM} transformed
to the gauge \eqref{ax} while $\delta{\cal W},\delta\Phi$ are of order $\omega$. 
Inserting this to Eqs.\eqref{eqs} and 
linearizing with respect to $\delta{\cal W},\delta\Phi$  
reveal that one can consistently choose 
\be                 \label{p1}
\delta {\cal W}=(-\omega+\tau_1\delta \psi^{1}_0+\tau_3\delta\psi^{3}_0)dt,~~~
\delta\Phi=0,
\ee
with 
$
\delta \psi^{1}_0+i\delta\psi^{3}_0=-e^{-i\vartheta}(H_{+}(r)\sin\vartheta
+iH_{-}(r)\cos\vartheta)/r.
$
The variables in the equations then separate,
\be                       \label{p3}
(\frac{d^2}{dr^2}
-\frac{w^2+1}{q_{\mp}r^2}-\frac{h^2}{2})H_{\pm}+
\frac{2w}{q_{\mp}r^2}\,H_{\mp}=\mp\omega\,\frac{ r h^2}{2}, 
\ee
with $q_{\pm}=2/(3\pm 1)$. 
For $\omega\neq 0$ the source term in these equations
forces  $H_{\pm}$ to be nonzero. Numerically solving these equations gives
a globally regular solution $H_{\pm}(r)$ (see Fig.\ref{fig1}) 
for which $H_{\pm}/r\to\pm\omega$ as $r\to\infty$, 
so that asymptotically $\delta\psi_0\to i\omega$ as it should. 
Passing back to the globally regular gauge \eqref{fields},
this perturbative solution describes a slow rotational excitation of the sphaleron. 
The angular momentum is obtained by linearizing $\int T^0_\varphi d^3{\bf x}$, 
\be                                  \label{JJ}
J={2\pi\, \omega}
\int_0^\infty h^2(1+\frac{1}{3r}\,H_{+}-\frac{2}{3r}\,H_{-})r^2dr,
\ee
which evaluates, {\it e.g.}, to $J=22.7\,\omega$ for $\beta=1$.

Summarizing, choosing a nonzero value of $\omega$ in Eqs.\eqref{fields}
breaks the spherical symmetry 
of the sphaleron down to the axial one, generates an electric field and 
produces an angular momentum. If $\omega$ is small then $J$ is small, 
the spinning configuration is only slightly nonspherical and can be 
perturbatively described by  Eqs.\eqref{p1}--\eqref{JJ}. For larger $\omega$
deviations from spherical symmetry become large and one needs to integrate 
the full system of partial differential equations (PDE's) 
\eqref{eqs} to construct the solutions. 
 
\begin{figure}[ht]
\hbox to\linewidth{\hss%
	\resizebox{7cm}{6cm}{\includegraphics{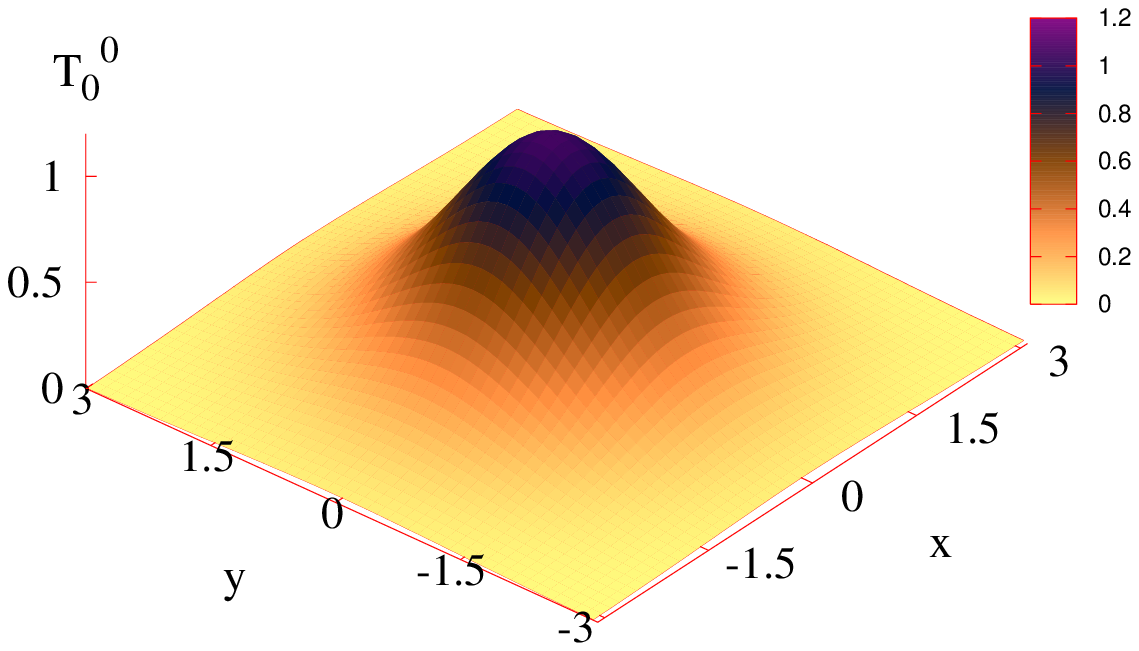}}
\hspace{5mm}%
        \resizebox{7cm}{6cm}{\includegraphics{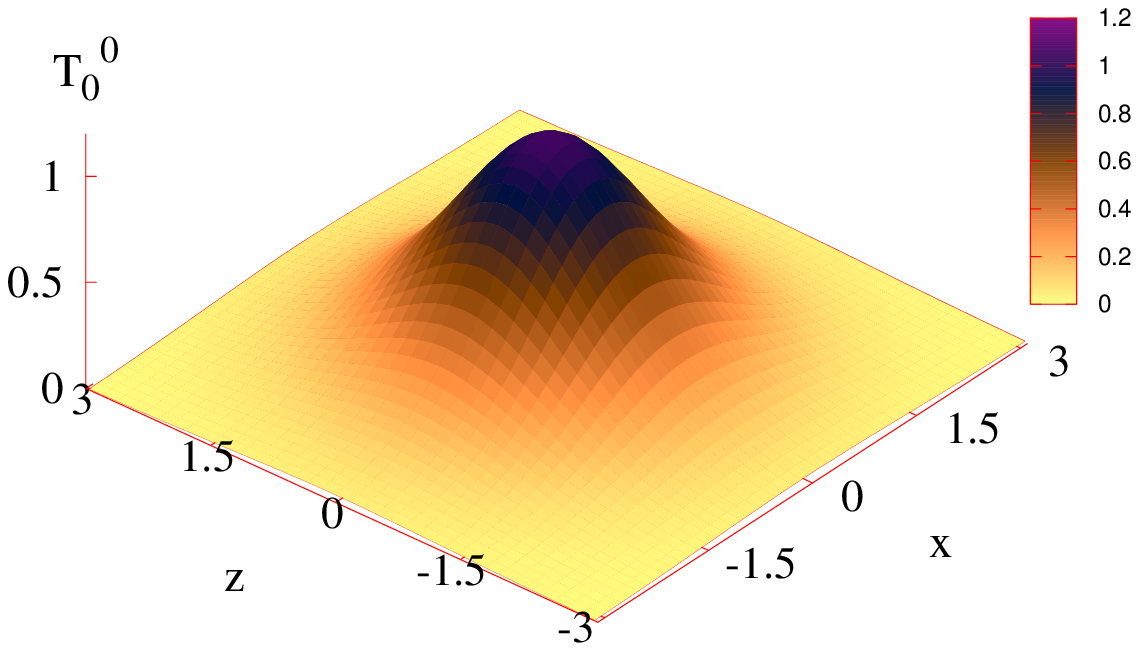}}	
\hss}
\hbox to\linewidth{\hss%
	\resizebox{7cm}{6cm}{\includegraphics{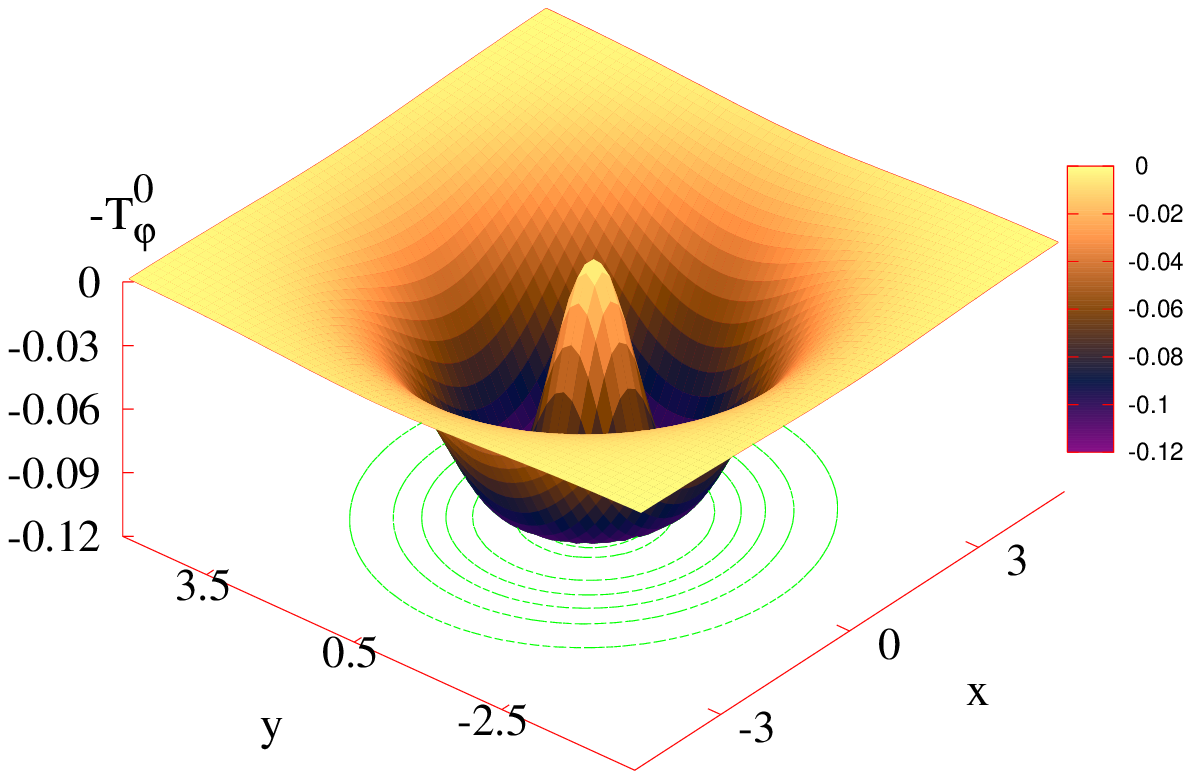}}
\hspace{5mm}%
        \resizebox{7cm}{6cm}{\includegraphics{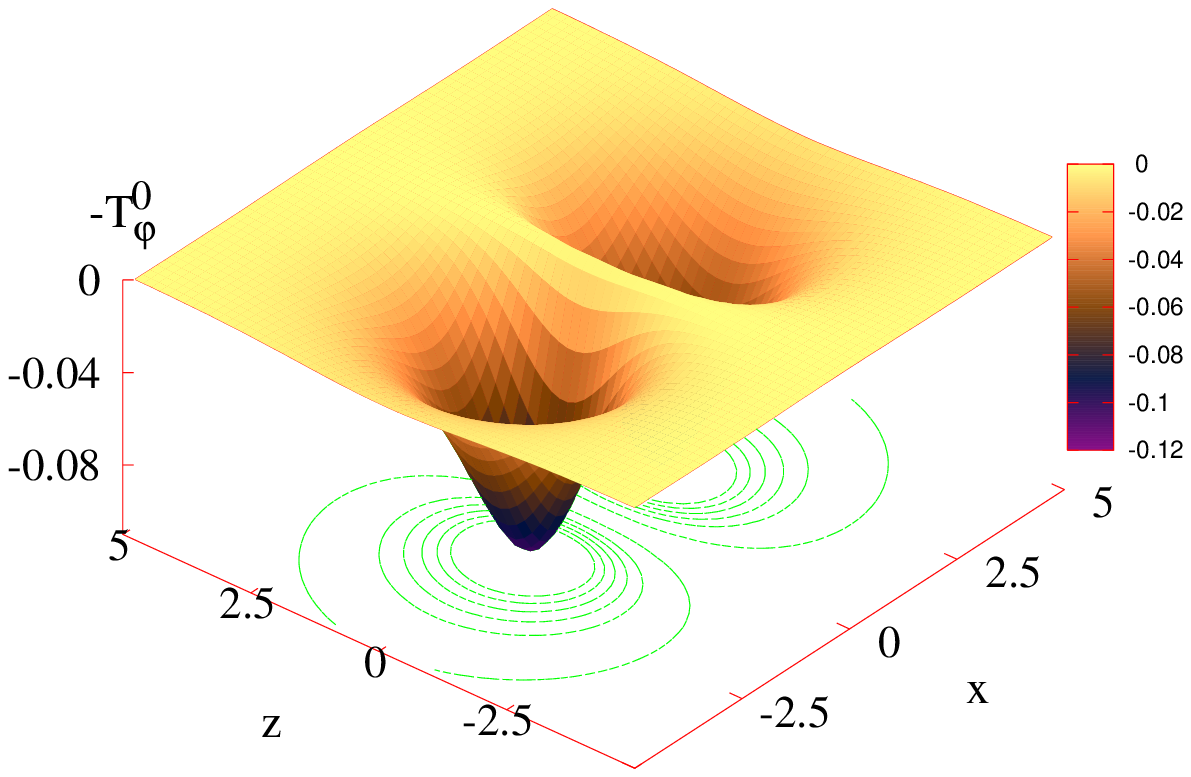}}	
\hss}

\caption{\small  
The energy density, $T_0^0$,  
and the negative (to better see the structure) 
angular momentum density, $-T^0_\varphi$, 
shown in the $z=0$ and $y=0$ planes
for
the spinning sphaleron with $\sin^2\thetw=0.23$, $\beta=2$, 
$\omega=0.433$. 
}
\label{fig3}
\end{figure}

We have performed our numerical calculations using the elliptic PDE solver FIDISOL 
 based  on the iterative Newton-Ralphson method \cite{FIDISOL}. We integrated a suitably 
discretized version of Eqs.\eqref{eqs} with the described above boundary conditions.   
Starting from the spherically symmetric sphaleron for $\thetw=\omega=0$ 
and increasing $\omega$ our numerics give nonperturbative, axially symmetric
solutions. The perturbative results are recovered for small $\omega$. However, 
as $\omega$ grows, deviations from the spherical symmetry, as well as 
$J$ and the energy $E$ increase, although 
the perturbative description seems to be still applicable as long as
$J/\omega\approx$const. (see Fig.\ref{fig3}, Fig.\ref{fig4}).   
It seems that there is a maximal value, $\omega_{\rm max}$, beyond
which no localized solutions exist.  Although it is difficult to approach this value
numerically, it appears  that $\omega_{\rm max}^2=g^2/2$, in which case the effective 
W-boson mass $\mw$ defined by Eq.\eqref{mass} vanishes, leading 
 to a delocalization
of the field configuration.

\noindent
\section{The case of $\thetw\neq 0$} 
\begin{figure}[ht]
\hbox to\linewidth{\hss%
	\resizebox{7cm}{6cm}{\includegraphics{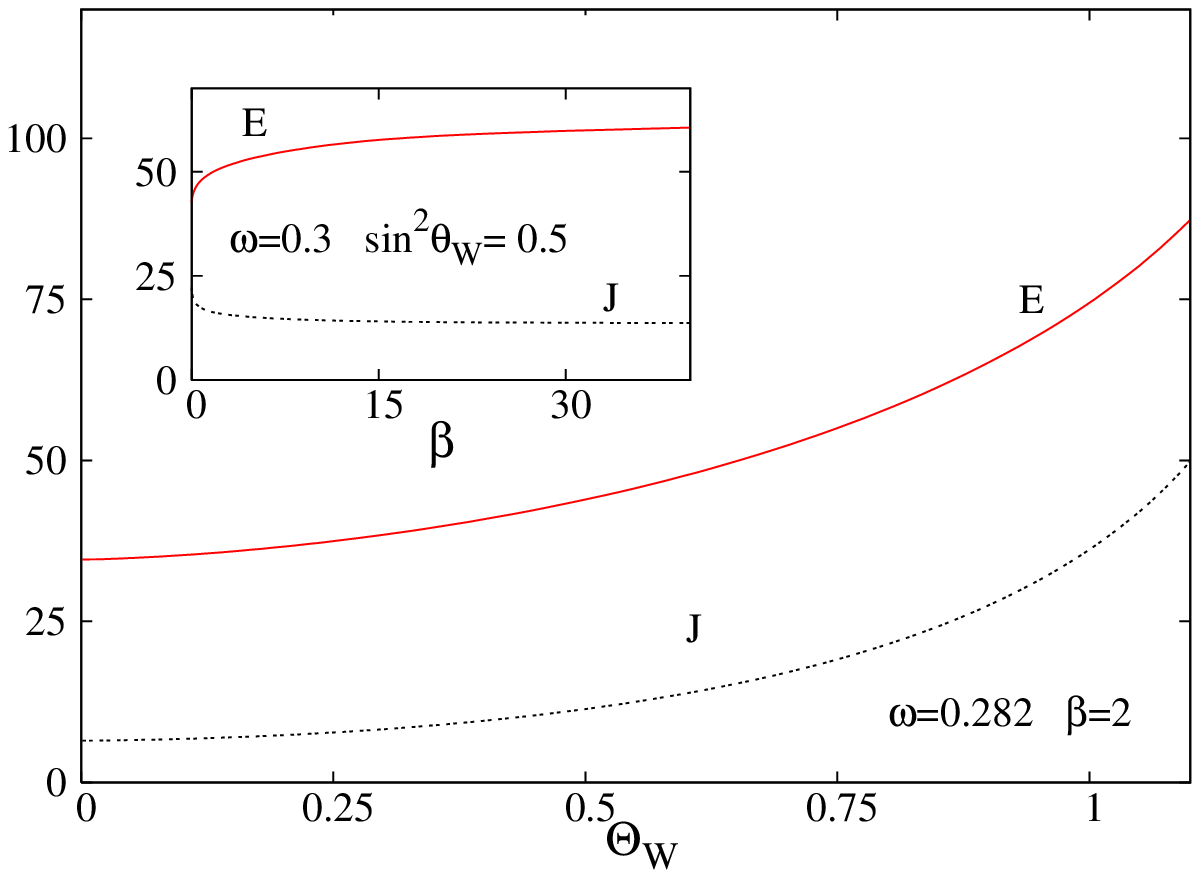}}
\hspace{5mm}%
        \resizebox{7cm}{6cm}{\includegraphics{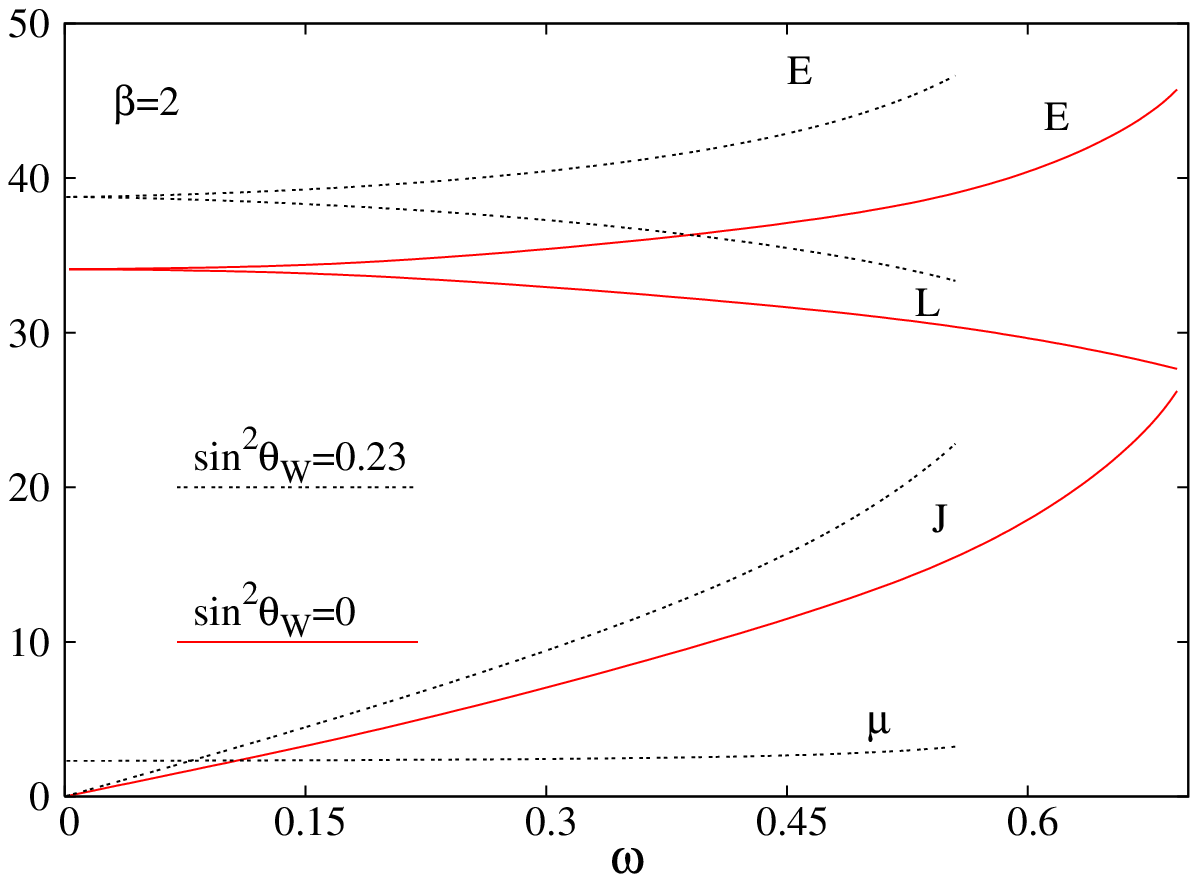}}	
\hss}

\caption{\small  Parameters of the spinning solutions.  
}
\label{fig4}
\end{figure}

In this case 
the described above features remain qualitatively the same, 
but the U(1) amplitudes $Y_a$ are  no longer constant 
and the solutions support a long-range electromagnetic field \eqref{Qmu} 
characterized by the electric charge $Q$ 
and magnetic dipole moments $\mu$.
If $\omega\to 0$ then 
$Q,J\to 0$ but $\mu$ remains finite (see Fig.\ref{fig4}), 
the solutions then becoming static and 
axially symmetric \cite{KK}.

Our numerics indicate that 
$J\approx C(\beta,\thetw)\,\omega E^2$ and so for $\omega\to\omega_{\rm max}$ 
one has $J\sim E^2$. A similar Regge-type behavior
 was predicted long ago by Nambu \cite{Nambu}  for the dumbbell --
monopole-antimonopole pair (MAP) connected by a Z-string segment and spinning around 
its center of mass. This suggests similarities with the dumbbell scenario, and the 
electric and magnetic current distributions for our solutions 
reveal indeed a MAP, but also 
an electric current loop encircling it, as schematically shown in Fig.\ref{fig1}.
Following \cite{HJ}
one can show that the MAP members
have magnetic charges $\pm 4\pi g^\prime/g$.
The current loop seems  to be stabilized by the MAP field, producing at the same 
time the Biot-Savart field that props the MAP up. The fields 
of the MAP and of the loop create together the sphaleron dipole moment \cite{HJ}, while for   
$\omega\neq 0$ there is also a momentum circulating along the loop
and creating the angular momentum $J$ directed {\it along} 
the MAP.  Therefore,
it is the loop that spins inside the sphaleron and not the MAP,
the whole system then resembling somewhat a vorton: vortex loop stabilized by the
centrifugal force \cite{DS}, \cite{RV}.  
This picture, however, can only be qualitative, since 
 the 
electromagnetic field is not uniquely defined off the Higgs vacuum \cite{Nambu}, \cite{HJ}.  

The static sphaleron is a saddle point solution  
relating to the top of the potential barrier between the topological vacua. 
The spinning sphalerons determine additional 
critical points of the action, and, by continuity, at least for small $J$, it is likely 
(but technically difficult to show) that 
these are also saddle points with one negative mode. Each of them 
presumably relates to the potential barrier separating 
the minimum energy states in sectors with fixed $Q=eJ$, 
as for example 
asymptotic states of $N$ spin-one W-bosons with the charge $Q=Ne$ and with $J=N$.  
The barrier transition amplitude 
is then determined by the sphaleron action density
$L=\int(-{\cal L})d^3{\bf x}$, which {\it decreases} with $J$ (see Fig.\ref{fig4}).  
The sphaleron-mediated transitions might therefore be enhanced in channels with nonzero
charge and angular momentum.

The fact that 
instead of just one saddle point of the action  
there are many of them increases the number
of transition channels. 
For example, one can argue that in hot electroweak plasma 
the topological transitions  in ZZ collisions, say, 
are mediated by the  $J=0$ sphaleron,
those in ZW$^{\pm}$ collisions -- by the $J=Q/e=\pm 1$ sphaleron, and so on. 
To get the total transition rate one should  sum over all channels, which may 
considerably affect the standard one-channel result \cite{RS}.
Of course, detailed calculations are necessary, since there could be competing effects,
as for example the Coulombian repulsion preventing the formation of charged sphalerons.
However, such a repulsion could perhaps be overcome by the weak force 
or by the plasma screening effects. In any case, the fact that there are many of them 
suggests that the overall contribution of the charged sphalerons may be important.

{We have checked that   
the multisphaleron, sphaleron-antisphaleron 
and vortex ring solutions  with $n\neq 1$ \cite{Kleihaus:1994yj}
also admit spinning 
generalizations. 
For $\thetw\neq 0$ they have $Q=neJ$. Charged
sphalerons  were also discussed
perturbatively \cite{Saffin}, and, 
after the preprint of the 
present paper was released,  
nonperturbatively in Ref.\cite{KKL}.  
 }

\noindent
\acknowledgements   
The work of E.R. was supported by ANR Grant NT05-$1_{-}$42856 
 `Knots and Vortons'. It is a pleasure to thank Gordon Semenoff for discussions.


\begin{thebibliography}{10}



\bibitem{KM}
  N.S.~Manton,
{{\XPEH Phys.Rev.}} {\bbf D28} (1983) 2019;
  F.R.~Klinkhamer and N.S.~Manton,
{{\XPEH Phys.Rev.}} {\bbf D30} (1984) 2212;
R.F.~Dashen, B.~Hasslacher and A. Neveu, 
{{\XPEH Phys.Rev.}} {\bbf D10} (1974) 4138.

\bibitem{KK}
  B.~Kleihaus, J.~Kunz and Y.~Brihaye,
{{\XPEH Phys.Lett.}} {\bbf B273} (1991) 100;
{{\XPEH Phys.Rev.}} {\bbf D46} (1992) 3587.
 

\bibitem{RS}
 V.A.~Rubakov and M.E.~Shaposhnikov,
{{\XPEH Phys.Usp.}} {\bbf 39} (1996) 461.

\bibitem{RV}
  E.~Radu and M.S.~Volkov, {\XPEH Phys.Rep.} {\bbf 468} (2008) 101.


\bibitem{VR}
  J.J.~Van der Bij and E.~Radu,
{{\XPEH Int.J.Mod.Phys.}} {\bbf A17} (2002) 1477;
{\it ibid} {\bbf A18} (2003) 2379.




\bibitem{VW}
  M.S.~Volkov and E.~Wohnert,
{{\XPEH Phys.Rev.}} {\bbf D67} (2003) 105006.



\bibitem{DS}
  R.L.~Davis and E.P.S.~Shellard, {\XPEH Nucl.Phys.} {\bbf B323} (1989) 209.






\bibitem{Nambu}
Y.~Nambu.
{{\XPEH Nucl.Phys.}} {\bbf B130} (1977) 505.


\bibitem{FM}
P.~Forgacs and N.~Manton.
{{\XPEH Comm.Math.Phys.}} {\bbf 72} (1980) 15.

\bibitem{RR}
C.Rebbi and P.~Rossi,
{{\XPEH Phys.Rev.}} {\bbf 22} (1980) 2010.



\bibitem{FIDISOL}
W. Sch\"onauer and R. Wei\ss, 
{{\XPEH J.Compt.Appl.Math.}} {\bbf 27} (1989) 279.

\bibitem{Kleihaus}
B.~Kleihaus.
{{\XPEH Phys.Rev.}} {\bbf D59} (1999) 125001.



\bibitem{HJ}
M.~Hindmarsh and M.~James,
{{\XPEH Phys.Rev.}} {\bbf D49} (1994) 6109. 



\bibitem{Kleihaus:1994yj}
  B.~Kleihaus and J.~Kunz,
{{\XPEH Phys.Lett.}} {\bbf B329} (1994) 61; 
  B.~Kleihaus, J.~Kunz and M.~Leissner,
{ {\XPEH Phys.Lett.}} {\bbf B663} (2008) 438.

\bibitem{Saffin}
  P.M.~Saffin and E.J.~Copeland,
{{\XPEH Phys.Rev.}} {\bbf D57} (1998) 5064.

\bibitem{KKL}
  B.~Kleihaus, J.~Kunz and M.~Leissner,
{\tt arXiv:0810.1142}


\end{thebibliography}
\end{document}